# Mixed dark matter with neutrino chemical potentials

Georg B. Larsen and Jes Madsen

*Theoretical Astrophysics Center and Institute of Physics and Astronomy, University of Aarhus,*

*DK-8000 Århus C, Denmark*

(February 9, 1995)

## Abstract

Mixed dark matter models with one low-mass (e.g. 2.4eV) neutrino flavor are shown to be in good agreement with observations if the neutrinos have non-zero chemical potentials. This agreement holds (except for the problem with a low-age Universe) even for high values of the Hubble-parameter. Massless neutrinos with non-zero chemical potentials may reconcile cold dark matter with observations. Some fine-tuning is required to avoid problems with Big Bang nucleosynthesis.

14.60.Pq, 95.35.+d, 98.65.-r, 98.80.-k

Typeset using REVTEX

The mixed dark matter (MDM) model has been very successful in reproducing the large scale structure observed in the Universe [1], and in fact is among the only models that can at the same time fit both the very large scales, as measured by COBE [2], and the studies of galaxy correlations, as recently summarized by Peacock and Dodds [3]. Very nice fits [4] are obtained if one neutrino of mass near 4.8eV (or even better, 2 neutrinos of mass near 2.4eV) is responsible for roughly 20% of the density in the Universe, with another 80% (apart from a minor fraction of baryons) in terms of an unknown form of cold dark matter (CDM). The relevance of such models is significantly strengthened if preliminary indications from oscillation experiments [5] of a 2.4eV muon neutrino mass are confirmed.

However, standard MDM models only fit the data if the expansion rate of the Universe has a definite value. In terms of the Hubble parameter, $H_0$, this value is $H_0 \approx 50$ kms$^{-1}$ Mpc$^{-1}$, or $h_0 \equiv H_0/100$ kms$^{-1}$ Mpc$^{-1} \approx 0.5$. Several recent investigations [6], while still marred by possible large systematic uncertainties, tend to give much higher values, in the range of $h_0 \approx 0.7$–0.8, which would be inconsistent with the MDM-models. (Such a high Hubble parameter would also make the Universe suspiciously young unless the cosmological constant is non-zero. An $\Omega = 1$ Universe has an age of only $t_0 = 6.5 h_0^{-1}$Gyr, giving $t_0 < 9.3$Gyr for $h_0 > 0.7$. Standard MDM-models are however in conflict with large scale structure data (regardless of the choice of $m_\nu$) even for $h_0 = 0.6$, where the age problem is not severe. In the present investigation we shall not dwell further on the age problem, but concentrate on reproducing the observed large scale structure).

The inconsistency between MDM and $h_0 > 0.5$ is related to the coupling between neutrino mass, neutrino density contribution, and neutrino free-streaming. As pointed out in the present investigation, mixed dark matter models can be reconciled with $h_0 > 0.5$ if this coupling changes, as in the case of a non-zero neutrino chemical potential [7]. Due to the preliminary experimental indications of a 2.4eV $\nu_\mu$ we shall focus in particular on a scenario where one such flavor is the hot component, but similar conclusions can be drawn for other choices of mass and number of flavors.

Neutrinos decoupling while ultra-relativistic are described by the momentum distribution

$$f(p) = \frac{1}{\exp\left(\frac{p-\mu}{T}\right) + 1}, \tag{1}$$

with $T = T_D a_D / a$, $\mu = \mu_D a_D / a$, where $T$, $a$, and $\mu$ denote temperature, scale factor, and chemical potential, and subscript $D$ indicates the value at decoupling.

In the detailed investigations described below we have integrated the full neutrino distribution function to get the density, free-streaming etc. Some approximate expressions in the ultra-relativistic and non-relativistic limits are useful, however.

Assuming neutrino-antineutrino equilibrium ($\bar{\mu} = -\mu$) the present neutrino contribution (including $\bar{\nu}$) to the cosmic density in units of the critical density is

$$\Omega_{\nu\bar{\nu}} = 2.382 \times 10^{-3} h_0^{-2} m_{2.4} \left[\eta^3 + \pi^2 \eta + 6F_2(-\eta)\right]. \tag{2}$$

In this relation $\eta \equiv \mu/T$ and $m_{2.4} \equiv m_\nu / 2.4$ eV. The neutrino is assumed to be single-handed. The function $F_2$ is

$$F_2(-\eta) = \Gamma(3) \sum_{r=0}^{\infty} (-1)^r \frac{e^{-(r+1)\eta}}{(r+1)^3}, \tag{3}$$

taking the value $3\Gamma(3)\zeta(3)/4$ for $\eta = 0$. Thus for non-degenerate neutrinos, $\Omega_{\nu\bar{\nu}} = 2.58 \times 10^{-2} h_0^{-2} m_{2.4}$, giving $\Omega_{\nu\bar{\nu}} \approx 0.21$ for a 4.8eV neutrino or two 2.4eV neutrinos and $h_0 = 0.5$.

In the ultra-relativistic regime a degenerate neutrino-antineutrino flavor contributes to the density like one non-degenerate flavor plus an additional

$$\Delta N_\nu = \frac{30}{7\pi^2}\eta^2 + \frac{15}{7\pi^4}\eta^4 \tag{4}$$

flavors. A problem with such a model is therefore the increase in relativistic density during Big Bang nucleosynthesis, leading to an overproduction of helium. To avoid this one needs for instance a slight degeneracy in electron neutrinos to push the weak interactions more in the direction of protons. Several authors have demonstrated that nucleosynthesis data only constrain the ratio $\eta_{\nu_e}/|\eta_{\nu_{\mu,\tau}}|$, and that the observations are in agreement with a large $\nu_\mu$ and/or $\nu_\tau$ chemical potential (see [8] for a review). In contrast to the cosmic net baryon number, which is known to be small, there are no strong observational or theoretical constraints on the lepton number, which is a sum of contributions from electrons and 3 neutrino flavors, where only the electron term is known to be small (from charge neutrality). The total lepton number could be large or it could be comparable to the baryon number. If it is comparable to the baryon number, this could occur because all neutrino terms were small, but it could also be due to cancellations. Even if a (GUT scale?) neutrino chemical potential is small, non-zero effective chemical potentials could appear later from particle decays [9]. While non-zero chemical potentials may seem like an *ad hoc* solution to the structure formation problem, it is not in conflict with any fundamental principles or observations.

We note in passing that a massless (or very low-mass) degenerate neutrino could be responsible for the increase in relativistic degrees of freedom needed in the scenario of Dodelson, Gyuk and Turner [10] to reconcile a pure CDM scenario with observations.

Finally it is worth noting that the mean momentum of degenerate neutrinos is higher than that of non-degenerate neutrinos ($3\mu/4$ in the limit $\eta \to \infty$ versus $3.151T$ for $\eta = 0$).

The relations above show the qualitative effects of neutrino degeneracy to be the following: A larger density contribution for fixed $m_\nu$, allowing a smaller neutrino mass to explain a given HDM-fraction for fixed $h_0$, or permitting a larger Hubble parameter for fixed $m_\nu$. At the same time, the free-streaming for fixed $m_\nu$ and $h_0$ increases somewhat. Detailed calculations summarized below show, that the final density perturbation spectrum is consistent with observations for a scenario with one degenerate 2.4eV neutrino (for instance the muon neutrino, assuming then an unstable $\nu_\tau$) and any choice of Hubble parameter in the interval $0.5 \leq h_0 \leq 1$. This in contrast to non-degenerate neutrinos, where $h_0 > 0.5$ is inconsistent with observations for any choice of neutrino mass.

We calculate the evolution of an initial density perturbation in a flat ($\Omega = 1$) Universe using linear theory and follow the approach by Bond and Szalay [11]. We assume one massive neutrino species with mass $m_\nu = 2.4$ eV and a non-zero chemical potential as the hot dark matter (HDM) component, leaving two massless neutrino species. The fraction of baryons in the Universe is only a few per cent, so to simplify calculations we neglect them completely when calculating the evolution of density perturbations. As our aim is to demonstrate the general features of a degenerate neutrino model rather than present detailed results for specific parameters the errors introduced thereby are not crucial. The CDM fraction is therefore just $\Omega_{\text{CDM}} = 1 - \Omega_{\nu\bar{\nu}}$.

We include the gravitational effect of photons in the tight-coupling approximation [11], neglecting photons after recombination. Only adiabatic density perturbations are considered, with initial conditions [11] $\delta_{\text{CDM}} = \frac{3}{4}\delta_\nu = \frac{3}{4}\delta_\gamma$. The scale factor $a$ is normalized such that $a = 1$ when the massive neutrinos become non-relativistic. We integrate the differential equations given in [11] (properly modified to allow degeneracy) from $a = 0.005$ and until $a = a_{\text{now}} \simeq 4000$. The scale factor today $a_{\text{now}}$ depends on the chemical potential since the mean momentum of the neutrinos grows with $\eta$. We assume $\eta > 0$ so that antineutrinos have $\bar{\eta} = -\eta < 0$. This means that the neutrinos will have a larger mean momentum than in the usual $\eta = 0$ case, and therefore more free-streaming so that more of the initial density perturbations will be erased, whereas the antineutrinos will free-stream less than usual. It is therefore practical to evolve the massive neutrino perturbations as two distinct perturbations $\delta_\nu$ and $\delta_{\bar{\nu}}$. For $\eta > 2$ neutrinos greatly outnumber antineutrinos.

The very large scale density perturbations, $\lambda > 500 h_0^{-1}$ Mpc enter the horizon well after the massive neutrinos have become non-relativistic. Thus the perturbations on these large scales are not damped by any free-streaming and just grow like $\delta \propto a^2$ when the Universe is radiation dominated, and as $\delta \propto a$ when the Universe shifts to be dominated by matter. The small scale perturbations enter the horizon when the massive neutrinos are still relativistic and therefore they are damped by free-streaming. The CDM perturbations grow at a lower rate than $\delta_{\text{CDM}} \propto a$ as long as the HDM perturbations $\delta_\nu$ and $\delta_{\bar{\nu}}$ are smaller than $\delta_{\text{CDM}}$; the more homogeneous HDM distribution suppresses the growth of CDM.

We calculate the transfer function $T(k)$ for ten wavenumbers ranging from $k = 4 \times 10^{-3} h_0 \text{Mpc}^{-1}$ to $k = 2 h_0 \text{Mpc}^{-1}$ (the corresponding wavelengths are $\lambda = 2\pi/k$; all wavelengths and wavenumbers are comoving with values given as they would appear to be today). The initial power spectrum is assumed to be of Harrison-Zel'dovich type as predicted by many theories of inflation. The power spectrum is given as $P(k) = AkT^2(k)$, where the normalization factor $A$ is determined by the quadrupole anisotropy of the microwave background radiation measured by COBE [2]. For this we have taken the value $Q = 17\mu K$. In order to compare directly with the large scale structure data extrapolated to the linear regime by Peacock and Dodds [3] we follow their notation and calculate the dimensionless power spectrum,

$$\Delta^2(k) \equiv \frac{k^3 P(k)}{2\pi^2} = \frac{A}{2\pi^2} k^4 T^2(k), \tag{5}$$

which we henceforth will refer to as the power spectrum. $\Delta^2(k)$ can be described as the contribution to the fractional density variance per logarithmic interval in $k$.

With the chemical potential as a free parameter, we calculate the power spectrum for a range of values for $h_0$ from 0.5 to 1.0 adjusting the value of $\eta$ for each value of $h_0$ in order to obtain a reasonable fit to the large scale structure data points [3]. In Figure 1 we show the calculated power spectrum for six models with values of $h_0$ ranging from 0.5 to 1.0; the corresponding values of $\eta$, $\Omega_{\nu\bar{\nu}}$ and shape parameter $\Gamma$ (to be defined later) are shown in Table 1.

The power spectrum $\Delta^2(k)$ is plotted as a function of the wavenumber in units of the Hubble parameter $h_0$. Almost the same shape of the power spectrum is obtained for the six different values of $h_0$.

If the density fluctuations are Gaussian (those generated by inflation are), all observable quantities of large scale structure can be directly calculated from the power spectrum as long as the perturbations are small, that is for only linear evolution. Thus if two power spectra are equal they would predict the same amount of large scale structure today. Within a few per cent the best-fitting models have $\eta = 4.7\,h_0 - 0.53$. (We note again, that even though we find very good agreement with the data for $0.5 < h_0 < 1$, the age problem probably forces $h_0 < 0.7$).

Figure 1 also shows the standard CDM model and the MDM power spectrum for one 2.4eV neutrino flavor with $\eta = 0$ and $h_0 = 0.5$. Both models are clearly inconsistent with the data. Increasing $h_0$ would make the curves bend the wrong way, making the fit even worse.

In order to further test if the calculated models can fit the properties of the real Universe, we derive some observable quantities from the power spectra. The *rms* mass fluctuation in a top-hat sphere of radius $r = 8h_0^{-1}$Mpc, $\sigma_8$, is found to be around 0.75 in accordance with other MDM models. From observations of galaxy surveys the *rms* fluctuation in galaxy number density is usually found to be $\sigma_{8,\text{gal}} = 1$, so galaxies are more clustered than mass, corresponding to a bias parameter of $b = \sigma_{8,\text{gal}}/\sigma_8 = 1.3$.

Another somewhat more direct measure of the mass distribution is the peculiar bulk velocity in spheres, which for a radius of $50h_0^{-1}$Mpc is observed to be $V_{50} = 335 \pm 80$km/s [4]. We find a model-value between 370km/s and 390km/s for $V_{50}$ in fine agreement with observations.

The number density of clusters of galaxies, $N_{\text{clust}}$, with a mass $M > 10^{15}h_0^{-1}M_\odot$ can be found using the Press-Schecter approximation [12]. We follow [4] and [13] using a Gaussian filter and the density contrast needed for collapse assumed to be $\delta_c = 1.5$. Observations give $N_{\text{clust}} = 4.0 \pm 2.0 \times 10^{-7}h_0^3$Mpc$^{-3}$ [4], whereas we get number densities from 6.4 to $10.4 \times 10^{-7}h_0^3$Mpc$^{-3}$ for the power spectra shown in Figure 1 when $h_0$ ranges from 1.0 to 0.5. Thus our models slightly overproduce clusters, particularly for $h_0 = 0.5$. Proper inclusion of baryons will reduce the power on cluster scales and below because initial density perturbations in baryons are not allowed to grow before recombination. This would reduce $N_{\text{clust}}$. It is also possible to reduce the power on large scales by choosing a lower normalization (there could be some gravitational wave contribution to the COBE anisotropy), and at the same time increase the power on small scales by choosing a smaller value of $\eta$.

The scale where the auto-correlation function crosses zero is observed to be larger than $40h_0^{-1}$Mpc [4]. We find a consistent value of $56h_0^{-1}$Mpc with very little variation. In general MDM models give a better fit than standard CDM, which predicts zero crossing already at $36h_0^{-1}$Mpc [4].

Thus MDM models with degenerate neutrinos fit the observed large scale structure rather well. On smaller scales of $\lambda \simeq 1h^{-1}$Mpc or less non-linear effects are crucial, and comparison with observations less reliable. One comparison which has been much discussed recently is with the observed number density of damped Lyman-alpha systems (DLAS), which constrain the mass fraction of collapsed baryons at high redshifts [14]. MDM models form the structure quite late and have some difficulty in forming enough collapsed structures to explain the DLAS [13,15]. From the observations which are quite uncertain at high redshifts, the mass fraction of collapsed baryons at $z = 3 - 3.5$ is determined to be $\Omega_{gas} = 6.0 \pm 2.0 \times 10^{-3}$. From the models $h_0 = 0.5$ and $h_0 = 1.0$ (the ones which differ most, particularly at smaller scales) we find values of $\Omega_{gas}$ to be at least $3.9 \times 10^{-3}$ and $6.5 \times 10^{-3}$ respectively, by following the methods described in [13]. (For the large wave numbers needed to calculate $\Omega_{gas}$, we have approximated the evolution of the power spectrum by the growth of the CDM component in a flat Universe with a uniform background density of neutrinos $\Omega_{\nu\bar{\nu}}$). The mass fraction of collapsed baryons was obtained assuming $\Omega_b = 0.075$ and assuming that all the gas is neutral, but some of the gas could be ionized or even removed by early star formation, thus the observed value of $\Omega_{gas}$ should here be taken as a lower limit. As we have not included baryons in the calculation of the power spectrum, the limits on $\Omega_{gas}$ should be taken only as an indication, that it is possible to make our MDM models with a chemical potential consistent also with the constraints from DLAS.

What is the dominant effect that modifies the initial power spectrum? The power spectrum is suppressed on small scales, because density perturbations inside the horizon can not grow before the Universe becomes matter dominated at time $t_{eq}$. Before this time all the neutrinos are relativistic (the massive neutrinos become non-relativistic just around $t_{eq}$). The shape of the power spectrum at $t_{eq}$ is roughly determined by the scale that crosses the horizon at that time, $\lambda_{eq}$. On scales larger than $\lambda_{eq}$, the transfer function $T(k) = 1$, whereas on scales smaller than $\lambda_{eq}$ the transfer function decreases roughly as $\lambda^2$ for pure CDM, and even stronger in MDM when neutrino perturbations are erased by free-streaming. The scale $\lambda_{eq}$ can be written as $\lambda_{eq} \simeq 10 \text{Mpc}(\Omega_0 h_0^2)^{-1}(g_*/3.36)^{1/2}$ [10], where $g_*$ is the effective number of particle degrees of freedom. The shape of the power spectrum is determined by $(h_0^{-1}\text{Mpc}/\lambda_{eq})$, so one can define a shape parameter, $\Gamma \equiv \Omega_0 h_0 (g_*/3.36)^{-1/2}$ [10]. In standard MDM models one has $\Omega_0 = 1$ and $h_0 = 0.5$ with three neutrino flavors giving $\Gamma = 0.5$. Stan-dard CDM can be made to fit the large scale structures observed by modifying the shape parameter $\Gamma$, the required value of $\Gamma \approx 0.25$ [3] can be obtained by adding more relativistic degrees of freedom increasing $g_*$ [10] (as noted earlier, a massless, degenerate neutrino might do exactly that).

In the MDM models shown in Figure 1, the increase of $\eta$ needed to fit the observed power spectrum for high values of $h_0$ increases the effective number of neutrino species in the relativistic era by $\Delta N_\nu$ given in Equation (4). This again increases $g_* = 3.36 + 0.454 \Delta N_\nu$ and makes the shape parameter $\Gamma$ stay almost constant at around 0.5 as needed for MDM models with $\Omega_{\nu\bar\nu} \simeq 0.25$; see Table 1.

The computed models are of course complex, and can only roughly be characterised by the shape parameter, in particular because neutrinos become non-relativistic very close to $t_{\text{eq}}$. The free-streaming properties also differ a bit. When $\eta$ is high the number density of massive neutrinos is higher although $\Omega_{\nu\bar\nu}$ stays roughly the same, and the mean momentum increases a little. Still, the dominant effect that modifies the power spectra, is the increase in radiation energy density delaying $t_{\text{eq}}$, which cancels out the increase in Hubble parameter.

We have assumed only one massive neutrino species, being the muon neutrino with a mass of 2.4 eV. There are of course other possibilities. Primack *et al.* [4] favor their C$\nu^2$DM model, which has 2 massive neutrino species (muon and tau) both with masses of 2.4 eV. That particular model makes a remarkably good fit to all the observed values of both large and small scale structures if $h_0 \approx 0.5$. It would be equally easy to make MDM models with two massive neutrinos and a medium/high value of the Hubble parameter fit the observed large scale structure by introducing a chemical potential for one or both of the neutrino species. Similarly for other values of the neutrino masses. Also, we have only considered a scenario where the massive $\nu_\mu$ was degenerate. The extra relativistic degrees of freedom could have been produced by a degenerate, massless neutrino as well.

To summarize, we have demonstrated how neutrino degeneracy makes MDM models viable for one low-mass neutrino flavor, even for medium/large values of the Hubble parameter. It would be necessary to perform more precise calculations of the power spectrum in order to determine exactly how well MDM models with a chemical potential can be made to fit all the observational constraints. No doubt this will be done if the neutrino mass detections are confirmed and the evidence for a large value of the Hubble parameter continues to increase.

## ACKNOWLEDGMENTS

This work was supported by the Theoretical Astrophysics Center under the Danish National Research Foundation, and the Theoretical Astroparticle Network funded by the European Human Capital and Mobility Program.

FIGURES

FIG. 1. The dimensionless power spectrum $\Delta^2(k)$ plotted as a function of wavenumber in units of $h_0 \mathrm{Mpc}^{-1}$. The MDM power spectra for six different choices of Hubble parameter ranging from 0.5 to 1.0 are shown, all normalized to the COBE quadrupole anisotropy. For each power spectrum we have chosen the value of $\eta$ such that a good fit to the observed linear power spectrum is obtained; the data points are taken from [3]. The corresponding values of $\eta$ are given in Table 1. Also shown for comparison is the standard CDM power spectrum, and the MDM power spectrum with $h_0 = 0.5$ and $\eta = 0$ for one 2.4eV neutrino flavor (solid curve close to CDM).

TABLES

TABLE I. Corresponding values of Hubble parameter, degeneracy, neutrino density fraction, and shape parameter, for the MDM models of Figure 1.

| $h_0$ | $\eta$ | $\Omega_{\nu\bar{\nu}}$ | $\Gamma$ |
|-------|--------|-------------------------|----------|
| 0.5 | 1.75 | 0.235 | 0.45 |
| 0.6 | 2.3 | 0.239 | 0.50 |
| 0.7 | 2.8 | 0.245 | 0.54 |
| 0.8 | 3.3 | 0.257 | 0.56 |
| 0.9 | 3.7 | 0.257 | 0.58 |
| 1.0 | 4.1 | 0.261 | 0.59 |

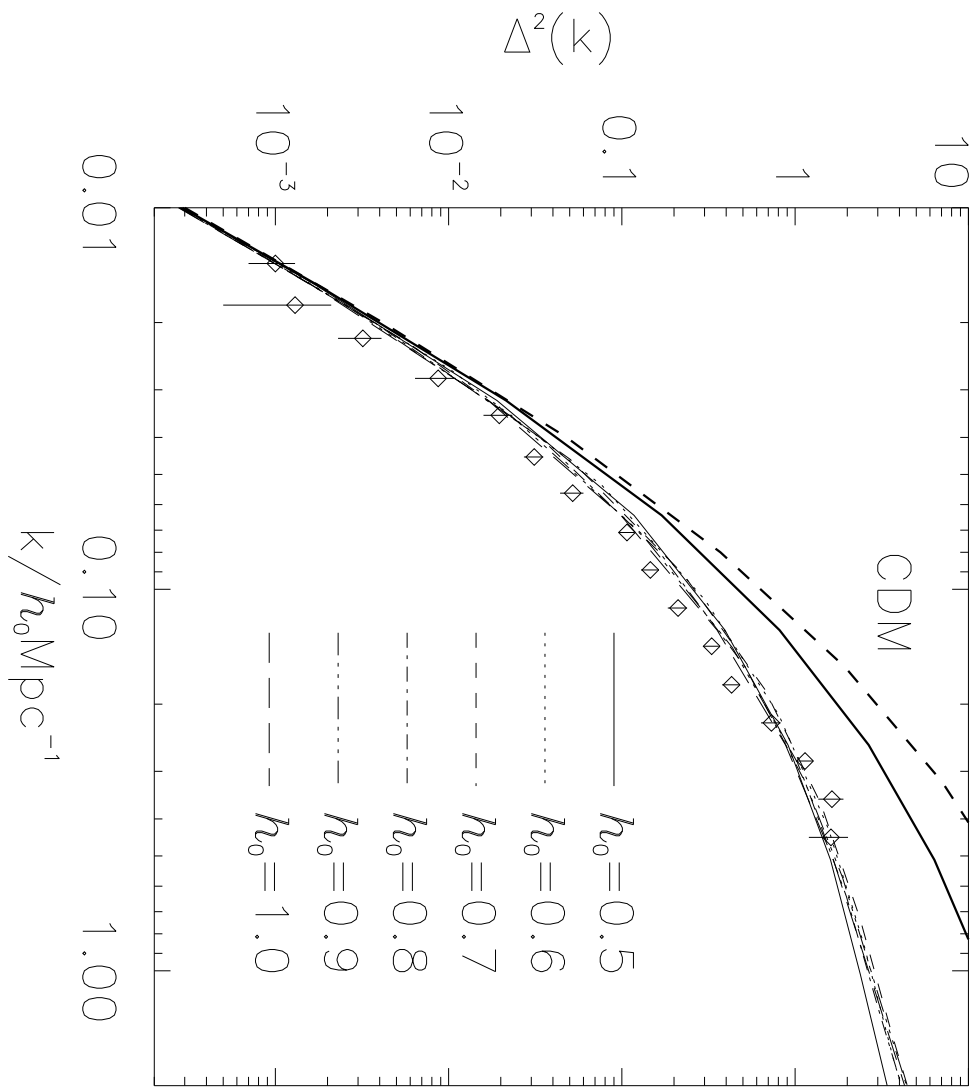